\documentclass[aps,pre,preprint,superscriptaddress,amsmath,amssymb,floatfix,showkeys]{revtex4}
\usepackage{bigints}
\usepackage{times}
\usepackage{natbib}
\usepackage{graphicx}
\usepackage{caption}
\usepackage{graphicx,amsmath,amssymb,amstext,amsfonts,color}
\usepackage{hyperref}
\usepackage{movie15}
\DeclareMathOperator{\bx}{\mathbf{x}}
\DeclareMathOperator{\by}{\mathbf{y}}

\begin{document}
\def\D{\Delta}
\def\d{\delta}
\def\r{\rho}
\def\p{\pi}
\def\a{\alpha}
\def\g{\gamma}
\def\ra{\rightarrow}
\def\s{\sigma}
\def\b{\beta}
\def\e{\epsilon}
\def\G{\Gamma}
\def\om{\omega}
\def\l{\lambda}
\def\f{\phi}
\def\w{\psi}
\def\m{\mu}
\def\t{\tau}
\def\c{\chi}
 \title{Individual-Based models for adaptive diversification in high-dimensional phenotype spaces}

\author{Iaroslav Ispolatov}
\email{jaros007@gmail.com}
\affiliation{
Departamento de Fisica, Universidad de Santiago de Chile,
Casilla 302, Correo 2, Santiago, Chile}

\author{Vaibhav Madhok}
\email{vmadhok@gmail.com}
\affiliation{Department of Zoology,  University of British Columbia, 6270 University Boulevard, Vancouver B.C. Canada, V6T 1Z4}

\author{Michael Doebeli}
\email{doebeli@zoology.ubc.ca}
\affiliation{Department of Zoology and Department of
  Mathematics,  University of British Columbia, 6270 University Boulevard, Vancouver B.C. Canada, V6T 1Z4}

\begin{abstract}
Most theories of evolutionary diversification are based on equilibrium assumptions: they are either based on optimality arguments involving static fitness landscapes, or they assume that populations first evolve to an equilibrium state before diversification occurs, as exemplified by the concept of evolutionary branching points in adaptive dynamics theory. Recent results indicate that adaptive dynamics may often not converge to equilibrium points and instead generate complicated trajectories if evolution takes place in high-dimensional phenotype spaces. Even though some analytical results on diversification in complex phenotype spaces are available, to study this problem in general we need to reconstruct individual-based models from the adaptive dynamics generating the non-equilibrium dynamics. Here we first provide a method to construct individual-based models such that they faithfully reproduce the given adaptive dynamics attractor without diversification. We then show that a propensity to diversify can by introduced by adding Gaussian competition terms that generate frequency dependence while still preserving the same adaptive dynamics. For sufficiently strong competition, the disruptive selection generated by frequency-dependence overcomes the directional evolution along the selection gradient and leads to diversification in phenotypic directions that are orthogonal to the selection gradient.
\end {abstract}


\maketitle



\section{Introduction}

Understanding the origin of biological diversity is one of the most fundamental problems in evolutionary biology. Traditional theories of diversification are based on static fitness landscapes and geographic isolation \cite{coyne_orr2004, gavrilets2004}. However, it has been realized that ecological interactions leading to frequency-dependent selection and dynamically changing fitness landscapes can generate adaptive diversification, a process that occurs without geographic isolation and requires ecological contact between the newly emerging species \cite{geritz_etal1998, dieckmann_doebeli1999, doebeli2011}.

There is a substantial amount of empirical evidence for adaptive diversification in sympatry \cite{rosenzweig_etal1994,rainey_travisano1998,rozen_lenski2000,berlocher_feder2002,friesen_etal2004,barluenga_etal2006,savolainen_etal2006,ryan_etal2007,legac_etal2012,herron_doebeli2013,plucain_etal2014}, and a plethora of different models have shown the theoretical feasibility of adaptive diversification (\cite{doebeli2011}, see also the comprehensive list of papers on Eva Kisdi's site mathstat.helsinki.fi/\textasciitilde kisdi/). In particular, the framework of adaptive dynamics has been extensively used in this context, because it is ideally suited to describe evolution on fitness landscapes that are changing dynamically due to frequency-dependent interactions, and to identify conditions that are conducive to adaptive diversification \cite{metz_etal1992,geritz_etal1998, doebeli2011}. In essence, adaptive dynamics is a formalism for deriving evolutionary trajectories in phenotype space. In most cases, this phenotype space is one-dimensional and represents scalar traits such as body size or other morphological or behavioural features. In this case, the conditions for adaptive diversification are very well understood and are based on the concept of evolutionary branching points. These are points in phenotype space with two essential properties: first, they are attractors for the evolutionary dynamics, and second, they are fitness minima. This may seem contradictory at first but becomes intuitively appealing once one takes into account that fitness landscapes are dynamic. Thus, as long as the evolutionary trajectory is away from the branching point, the fitness landscape determining the trajectory has a minimum so that the current phenotypic position of the evolving population is on one side of this minimum. As the population climbs away from the fitness minimum, the fitness landscape changes in such a way that the minimum eventually catches up with the trajectory, at which point the dynamics has reached its equilibrium, while the population sits on a fitness minimum. Because of this, phenotypic mutants on either side of the current position of the population can invade, resulting in adaptive diversification. For example, if this scenario is modelled with individual-based models, in which a population is represented as a cloud of points in phenotype space, then this cloud first converges towards the position in phenotype space corresponding to the evolutionary branching point, and subsequently splits into two separate clouds that diverge from each other, thus yielding and elementary representation of  evolutionary diversification. The existence of evolutionary branching points is a robust feature of adaptive dynamics in low-dimensional phenotype spaces and has been demonstrated in many different settings \cite{doebeli2011}.

In the majority of models studied to date, adaptive diversification occurs according to the scheme just described: first the evolutionary trajectory converges to a branching point, and then the population splits into diverging and coexisting phenotypic branches. For example, it has been shown that, conditional on convergence to an evolutionary equilibrium, the likelihood of evolutionary branching increases with the dimension of phenotype space \cite{doebeli_ispolatov2010,svardal_etal2014,debarre_etal2014}. However, it has recently also been shown that in high-dimensional phenotype spaces, the adaptive dynamics may never converge to an equilibrium in the first place \cite{doebeli_ispolatov2014}. In fact, in a general class of competition models, the likelihood that adaptive dynamics are chaotic approaches 1 if the dimension of phenotype space is high enough \cite{doebeli_ispolatov2014}. In particular, such dynamics would never converge to an evolutionary branching point. A resolution of this apparent paradox has been found in  \cite{ito_dieckmann2014}, which shows analytically that sufficiently strong disruptive selection can ``overcome'' a  small but non-zero selection gradient and produce sustained diversification in the direction perpendicular to the selection gradient. The analysis in \cite{ito_dieckmann2014} is local in the sense that it describes the initial phase of evolutionary branching when the resident is moving along an assumed and constant gradient in one phenotypic direction. Here we study the question of diversification in non-stationary resident populations when the selection gradient along which the resident is evolving is not assumed, but is a result of complex interactions in high-dimensional phenotype space. In particular, the trajectory of the resident is not simply given by a constant gradient, but can exhibit more complicated dynamics, such as cycles and chaos. 

In the following, we first briefly recall how the conditions for diversification of \cite{ito_dieckmann2014} apply in scenarios with complicated evolutionary dynamics of the resident. To study the full dynamics of diversification in such scenarios, we then reconstruct individual-based  models from an adaptive dynamics model that generates complicated trajectories. It is known \cite{dieckmann_law1996, champagnat_etal2006, champagnat_etal2008} that adaptive dynamics is a particular large-population limit of underlying stochastic, individual-based processes. What seems to be less appreciated is that there are in fact very many different individual-based models that yield the same adaptive dynamics. However, these different individual-based models differ in their propensity to diversify. Here we show that given an adaptive dynamics model, there is a particular ``minimal'' individual-based model that never diversifies and that produces qualitatively the same trajectories as the given adaptive dynamics model. We then show how, in light of the conditions for diversification in \cite{ito_dieckmann2014}, this minimal model can be altered to generate adaptive diversification, yet still have the same adaptive dynamics limit. Similar results hold for partial differential equation (PDE) models, which represent an intermediate limit between the individual-based and the adaptive dynamics \cite{champagnat_etal2006, champagnat_etal2008}. Our work appears to be one of the first attempts to  study non-equilibrium evolutionary dynamics  of complex, high-dimensional phenotypes using  individual-based and PDE models.

 \section{Adaptive dynamics and diversification}
 
 As in \cite{doebeli_ispolatov2014}, we study a general class of models for frequency-dependent competition in which ecological interactions are defined by $d$-dimensional phenotypes, where $d\geq1$. For example, one can imagine that a high-dimensional phenotype of individuals is given by the efficiencies of several metabolic pathways, or my multiple morphological characteristics. The ecological interactions are determined by a competition kernel $\alpha(\mathbf{x}, \mathbf{y})$ and by a carrying capacity $K(\mathbf{x})$, where $\mathbf x,\mathbf y\in\mathbb{R}^d$ are the $d$-dimensional phenotypes of competing individuals. The competition kernel $\alpha$ measures the  competitive impact that an individual of phenotype $\mathbf x$ has on an individual of phenotype $\mathbf y$, and in the sequel we always assume that  $\alpha(\mathbf{x}, \mathbf{x})=1$ for all $\mathbf x$. Assuming logistic ecological dynamics, $K(\mathbf{x})$ is then the equilibrium density of a population that is monomorphic for phenotype $\mathbf x$. The per capita growth rate of a rare mutant phenotype $\mathbf y$ in the resident $\mathbf x$ is then given by 
 \begin{align}
 \label{fitness}
f(\mathbf{x}, \mathbf{y}) = 1 - \frac{ \alpha(\mathbf{x}, \mathbf{y}) K(\mathbf{x})}{K(\mathbf{y})}
\end{align}
(see \cite{doebeli2011,doebeli_ispolatov2014} for more details).  The function $f(\mathbf{x}, \mathbf{y}) $ is the invasion fitness from which the adaptive dynamics is derived based on the selection gradients
 \begin{align}
\label{sg}
s_i(\mathbf{x}) \equiv \left. \frac{\partial f(\mathbf{x}, \mathbf{y})}{\partial y_i}\right|_{ \mathbf{y}=\mathbf{x}} =  - \left. \frac{\partial \alpha(\mathbf{x}, \mathbf{y})}{\partial y_i}\right|_{ \mathbf{y}=\mathbf{x}} + \frac{\partial K(\mathbf{x})}{\partial x_i}\frac{1}{K(\mathbf{x})},
\end{align}
The selection gradients define a system of differential equations on phenotype space $\mathbb{R}^d$, which is given by
 \begin{align}
 \label{AD1}
 \frac{d\mathbf{x}}{dt} = \mathbf M\cdot\mathbf{s}(\mathbf{x}).
\end{align} 
Here $\mathbf M$ is the mutational variance-covariance matrix, which generally influences the speed and direction of evolution. For simplicity, we assume here that this matrix is the unit matrix. For more details on the derivation of the adaptive dynamics (\ref{AD1}) we refer to a large body of primary literature (e.g. \cite{dieckmann_law1996,geritz_etal1998, diekmann2003, leimar2009, doebeli2011}). We note that the adaptive dynamics (\ref{AD1}) can be derived analytically as a large-population limit  of an underlying stochastic, individual-based model that is again defined by the competition kernel $\alpha(\mathbf{x}, \mathbf{y})$ and the carrying capacity $K(\mathbf{x})$ \cite{dieckmann_law1996, champagnat_etal2006,champagnat_etal2008}.

The system of ODEs  (\ref{AD1}) describes the trajectory of an evolving monomorphic population in phenotype space $\mathbb{R}^d$. Typically, the goal of an adaptive dynamics analysis is to find equilibrium attractors for these trajectories, which are called singular points. These are points $\mathbf x^*$ in phenotype space at which the selection gradient vanishes, $\mathbf{s}(\mathbf{x}^*)=0$. However, in general the trajectories may also exhibit more complicated dynamics, such as limit cycles or chaos. In fact, \cite{doebeli_ispolatov2014} have argued that in high-dimensional phenotype spaces, almost all systems of the form  (\ref{AD1}) will be chaotic.

For scalar traits, i.e., when the dimension of phenotype space is one, the adaptive dynamics typically converges to a singular point $x^*$, and it is well known that these points are often evolutionary branching points, i.e., starting points for adaptive diversification. This can be seen by expanding the invasion fitness function with respect to the mutant coordinate $y$ to second order,
\begin{align}
 \label{TEF}
 f(x,y) = &f(x, x) + \left.\frac {\partial f(x, y)}{\partial{y}}\right|_{y=x} (y-x)+  \left. \frac{\partial^2 f(x, y)}{\partial y^2} \right|_{ y=x}\frac{(y-x)^2}{2}.
\end{align}
The first term on the right hand side, $f(x, x)$, is zero for all $x$ since the resident population is assumed to be at the ecological equilibrium. The coefficient of the linear term is the selection gradient, which vanishes at a singular point by definition. In the neighbourhood of singular points, $  {\partial f(x, y)}/{\partial{y}}|_{ y=x} \rightarrow 0$, and the second order term in eq. (\ref{TEF}) becomes significant. In particular, if the singular point is a maximum of the invasion fitness, i. e. ${\partial^2 f(x^*, y)}/{\partial y^2}|_{ y=x^*} < 0$, no nearby mutants can invade and the population remains monomorphic. However, when the invasion fitness has a minimum at the singular point, i. e. when  ${\partial^2 f(x^*, y)}/{\partial y^2}|_{ y=x^*} > 0$, the stationary point  is an evolutionary branching point. Evolutionary branching has been studied extensively for scalar traits (\cite{metz_etal1992, geritz_etal1998, doebeli_dieckmann2000, doebeli2011}). 

An extension of these arguments to high-dimensional phenotype spaces reveals two interesting trends. On the one hand, the conditions for evolutionary branching at a singular point, i.e., at an equilibrium of the adaptive dynamics, are progressively easier to satisfy for increasing dimensionality \cite{doebeli_ispolatov2010,debarre_etal2014, svardal_etal2014}. On the other hand, and as mentioned before, the fraction of trajectories that converge to a singular point decreases  and essentially vanishes for very high-dimensional phenotypes, as more and more trajectories become chaotic \cite{doebeli_ispolatov2014}.  Thus, adaptive diversification by means of convergence to evolutionary branching points and subsequent diversification becomes less likely in high-dimensional phenotype spaces, and the possibility of non-equilibrium adaptive dynamics makes it necessary to study diversification not just from evolutionary branching points, but more generally from any point on an evolutionary trajectory.  Indeed, \cite{ito_dieckmann2014} have studied this problem analytically under the assumption that the resident is under directional selection in a particular phenotypic direction.

To see how the analysis of \cite{ito_dieckmann2014} applies in cases with complicated evolutionary trajectories, consider the multidimensional generalization of the expansion of the invasion fitness  (\ref{TEF}),
\begin{align}
 \label{HTEF}
 f(\mathbf{x},\mathbf{y}) = &f(\mathbf{x}, \mathbf{x}) +\left(\frac {\partial f(\mathbf{x}, \mathbf{y})}{\partial{y_i}}, ..., \frac {\partial f(\mathbf{x}, \mathbf{y})}{\partial{y_d}}\right)_{\mathbf{y=x}}  (\mathbf{y}-\mathbf{x}) +\frac{1}{2} (\mathbf{y}-\mathbf{x})^T \mathbf{H\mathbf(x)} (\mathbf{y}-\mathbf{x}),
\end{align}
where $\mathbf{H({x})}$ is the Hessian matrix of second derivatives of the invasion fitness function with respect to the mutant trait value $\mathbf y$, evaluated at the resident trait value $\mathbf x$:
\begin{align}
\mathbf{H}(\mathbf{x}) = \begin{pmatrix} \left. \frac{\partial^2}{\partial y_1 \partial y_1} f(\mathbf{x}, \mathbf{y})\right|_{\mathbf{y}=\mathbf{x}} & ... & \left. \frac{\partial^2}{\partial y_1 \partial y_d} f(\mathbf{x}, \mathbf{y})\right|_{\mathbf{y}=\mathbf{x}} \\...&...&...\\ \left.\frac{\partial^2}{\partial y_d \partial y_1} f(\mathbf{x}, \mathbf{y})\right|_{ \mathbf{y}=\mathbf{x}}&...&\left. \frac{\partial^2}{\partial y_d \partial y_d} f(\mathbf{x}, \mathbf{y})\right|_{\mathbf{y}=\mathbf{x}}\end{pmatrix}
\end{align}
 As in the one-dimensional case, the selection gradient, which determines the evolutionary trajectory, is the linear term in (\ref{HTEF}). However, the linear term is not relevant for the remaining $d-1$ dimensions that are orthogonal to the selection gradient. In this orthogonal subspace, the quadratic terms determine whether invasion is possible. Specifically, the interplay between the eigenvalues of the  Hessian matrix $ \mathbf{H}(\mathbf{x})$ and the selection gradient determine the curvature of the invasion fitness in the directions that are orthogonal to the direction in which the population is evolving \cite{ito_dieckmann2014}. In particular,  if $ \mathbf{H}(\mathbf{x})$ is negative definite, then the fitness function, viewed as a function on the subspace that is orthogonal to the selection gradient, has a maximum at the current resident, and only mutant phenotypes that have a component in the direction of the selection gradient can invade.  Thus, in this case no diversification is expected, and instead the evolving population simply follows the trajectory determined by the selection gradient. However, if the Hessian matrix has an eigenvector that has a positive eigenvalue and a non-zero projection onto the orthogonal subspace, then there are directions in phenotype space that are orthogonal to the selection gradient and along which the diversification is possible. The precise conditions for diversification in terms of the curvature of the invasion fitness in the direction orthogonal to the selection gradient and the magnitude of the selection gradient have been investigated in \cite{ito_dieckmann2014}. Roughly speaking, diversification occurs if the eigenvalues of the orthogonal Hessian are positive enough. 
 
It is important to note that if the population evolves on a non-equilibrium trajectory due to the first order term in (\ref{HTEF}), then the direction of the selection gradient continually changes, and hence so does the subspace that is orthogonal to the selection gradient. In particular, phenotypic directions along which the invasion fitness has a minimum may change over time, which may impede diversification, as the conditions for diversification may be satisfied only temporary. Nevertheless, it seems clear that in principle, the rather restrictive conditions for adaptive diversification in scalar traits, requiring the adaptive dynamics to converge to a branching point, can easily be circumvented in high-dimensional phenotype spaces, in which diversification can occur in directions that are orthogonal to the current direction of evolution. Below we use individual-based models to confirm that the essential positivity of the local curvature of the Hessian of the invasion fitness is indeed necessary and sufficient for  adaptive diversification during non-equilibrium evolutionary dynamics.

\section{Reconstructing individual-based models with different patterns of diversification} 
 
Our main goal is to construct individual-based models that correspond to a given adaptive dynamics (\ref{AD1}). In individual-based models, a population is represented by a ``cloud'' of points that moves through phenotype space over time. The points in the cloud are the individuals, each of which is assigned a birth rate and a death rate at any given point in time. The evolutionary dynamics unfolds through an algorithm according to which individuals with higher birth rates are more likely to give birth, and individuals with higher death rates are more likely to die. For birth events, the simplest assumption is asexual reproduction, and when individuals give birth, the offspring has a phenotype that is similar to the phenotype of the parent. There are various ways of implementing such algorithms, and the one used below is described in the Supplementary Online Materials (SOM). We refer to the literature for a more detailed description of such algorithms (e.g. \cite{doebeli2011, champagnat_etal2006, champagnat_etal2008}). In such individual-based models, a monomorphic population corresponds to a single compact cloud, and diversification occurs when an evolving population splits into two clusters in phenotype space. Thus,  after diversification, there are multiple coexisting clouds of points in phenotype space, each representing a descendent lineage.  
 
 An alternative way to describe the evolving population is by using partial differential equations (PDE's) for the dynamics of population distributions in phenotype space (see SOM). In that case, diversification corresponds to the formation of multiple modes in the evolving distribution. in fact, these deterministic models are themselves large-populations limits of the individual-based models, but they are more general than the adaptive dynamics limit in the sense that fewer assumptions are necessary to derive the PDE limit. This has been nicely described in \cite{champagnat_etal2008}. Due to computational limitations of numerical solutions of PDE's, solving the PDE models is currently only feasible in relatively low-dimensional phenotype spaces. 

When constructing either-individual based models or PDE models that correspond to a given adaptive dynamics (\ref{AD1}), a fundamental problem arises. The adaptive dynamics (\ref{AD1}) is determined by the selection gradient (\ref{sg}), which is in turn determined by the derivatives of the invasion fitness function (\ref{fitness}), and hence by the derivatives of the competition kernel and the carrying capacity. Because taking derivatives does obviously not yield a one-to-one correspondence, there are infinitely many  invasion fitness functions with different competition kernels and carrying capacities that have the same derivatives, and hence yield the same adaptive dynamics. Since both individual-based models and PDE models require knowledge of the full competition kernel and carrying capacity (see SOM), this implies that there are infinitely many different such models that correspond to the same adaptive dynamics (\ref{AD1}). Importantly, while yielding the same adaptive dynamics, the different competition kernels and carrying
capacities produce different patterns of diversification.

\subsection{The ``minimal" individual-based model}

To illustrate the above points, we reconstruct individual-based models from a given adaptive dynamics model by writing the adaptive dynamics equations (\ref{sg}, \ref{AD1}) as 
\begin{equation}
\label{Gen_AD}
\frac{dx_i}{dt} =  - \frac{\partial \alpha(\mathbf{x}, \mathbf{y})}{\partial y_i}_{\big | \mathbf{y}=\mathbf{x}} + \frac{\partial \ln[K(\mathbf{x})]}{\partial x_i} \equiv w_i(\mathbf{x}) + u_i(\mathbf{x}),
\end{equation}  
where we denote by  $w_i(\mathbf{x})$ and $u_i(\mathbf{x})$ the terms coming from the competition kernel and the carrying capacity, respectively. 

We first note that the part of the selection gradient that is due to the carrying capacity function (i.e., the $u_i(\mathbf{x})$ terms) essentially generate a hill climbing process on the function $K(\mathbf{x})$. As in \cite{doebeli_ispolatov2014} we assume that this hill-climbing process is of the simplest possible form by assuming that the carrying capacity function is given by the unimodal function 
\begin{align}
 \label{K}
 K(\mathbf{x})=K_0\exp\left(-\sum_i x_i^4/4\right).
\end{align}
This implies that the terms $u_i(\mathbf{x})=-x_i^3$ correspond to a stabilizing component of selection, with $\mathbf{x}=0$ being the optimal phenotype, because this phenotype corresponds to the highest carrying capacity.

Under these assumptions, we first show that regardless of the specific form of the $w_i(\mathbf{x})$-terms in (\ref{Gen_AD}), there is a ``minimal" competition kernel for which the  Hessian matrix of the invasion fitness is always negative definite and hence the evolving population should remain monomorphic in the individual-based model.

Specifically, we define the minimal competition kernel as
\begin{align}
 \label{ACCC}
\alpha({\mathbf{x}, \mathbf{y}}) = \exp\Big(\sum_{i=1}^{d} w_i(\mathbf{x})(x_i - y_i)\Big),
\end{align}
where the $w_i(\mathbf{x})$ are taken from the adaptive dynamics (\ref{Gen_AD}). It is easy to see that the adaptive dynamics corresponding to this competition kernel, as well as to the carrying capacity (\ref{K}), is given by (\ref{Gen_AD}) as desired. 
Moreover, with this competition kernel, we calculate the elements of the Hessian (7) of the invasion fitness function (2) as
\begin{align}
 \label{FHE}
\mathbf H(\mathbf{x})_{ij} = &- w_i(\mathbf{x}) w_j(\mathbf{x}) + w_i(\mathbf{x}) u_j(\mathbf{x}) +w_j(\mathbf{x}) u_i(\mathbf{x}) - u_i(\mathbf{x})u_j(\mathbf{x})  +\frac{1}{K(\mathbf{x})} \frac{\partial^2 K({\mathbf{x}})}{\partial x_i \partial x_j}
\end{align}
This matrix and can be written more succinctly as 
\begin{align}
 \label{FHEM}
\mathbf{H}(\mathbf{x}) = &- \begin{pmatrix}  w_1 - u_1\\...  \\w_d - u_d\end{pmatrix} \begin{pmatrix}  w_1 - u_1, & ...,& w_d - u_d\end{pmatrix} + \frac{\mathbf{H_K}(\mathbf{x})}{K(\mathbf{x})},
\end{align}
where $\mathbf{H_K}(\mathbf{x})$ is the Hessian of the carrying capacity, which is by assumption negative definite (and where we have omitted the argument $\mathbf{x}$ from the terms $w_i(\mathbf{x})$ and $u_i(\mathbf{x})$).

The first term in (\ref{FHEM}) is a rank 1 matrix 
and its only non-zero eigenvalue $\l$ is negative:
\begin{align}
 \label{lambda}
\l = -\sum_i^d [w_i(\mathbf{x}) - u_i(\mathbf{x})]^2.
\end{align}

As consequence, with the minimal competition kernel the Hessian of the invasion fitness function minimal competition kernel is always negative definite, independent of the current resident phenotype, and hence diversification is not expected anywhere along the trajectory of the adaptive dynamics. 

As an example, consider the adaptive dynamics studied in \cite{doebeli_ispolatov2014},
\begin{align}
\label{main1}
\frac{dx_i}{dt}=\sum_{j=1}^d b_{ij} x_j + \sum_{j,k=1}^d
a_{ijk}x_j x_k - x_i^3, \; i=1,\ldots,d.
\end{align} 
where $b_{ij}$ and $a_{ijk}$ are arbitrary  coefficients. The corresponding minimal competition kernel is derived from these coefficients as 
\begin{align}
 \label{CCC}
\alpha({\mathbf{x}, \mathbf{y}}) = \exp\left[\sum_{i,j=1}^{d} b_{ij}x_{j}(x_i - y_i) + \sum_{i,j,k=1}^{d} a_{ijk} x_jx_k(x_i - y_i)\right],
\end{align}
while the carrying capacity is given by (\ref{K}) as before. The prediction is that no matter what adaptive dynamics the coefficients $b_{ij}$ and $a_{ijk}$ generate, the corresponding individual-based model obtained from the competition kernel (\ref{CCC}) and the carrying capacity (\ref{K}) do not show diversification and produce clouds of points that move roughly along the same trajectories as the adaptive dynamics model (\ref{main1}). 

Our extensive simulations of individual-based models for many different choices of the coefficients $b_{ij}$ and $a_{ijk}$ support this prediction. Examples are shown in Figures 1-3 for two salient non-equilibrium behaviours of the adaptive dynamics:  a periodic attractor and a chaotic attractor. In both cases, the figures show the comparison of the adaptive dynamics (\ref{main1}) with the corresponding individual-based model. Even though the individual-based simulation do of course not remain strictly monomorphic, they also do not break up into distinct clusters. Instead the phenotypic variance remains constrained to a single cohesive cluster, and the trajectory of the center of mass of this evolving cluster follows the adaptive dynamics trajectory. Naturally, due to stochasticity of the individual-based solution, the center of mass trajectory is inevitably noisy. Nevertheless, the size, shape, and the direction of motion of the individual-based trajectory correspond to those of the adaptive dynamics solution.   For chaotic adaptive dynamics, this is a notable finding, as it was not at all clear a priori whether it is possible to find individual-based models that follow a chaotic trajectory without loss of cohesiveness and concomitant increase of phenotypic variance. 

A similar comparison can be performed for the adaptive dynamics and the corresponding PDE model, although this is currently only feasible for dimensions $d\leq3$. In such cases, the results are the same: the PDE model with the minimal competition kernel does not diversify, i.e., the phenotype distribution remains unimodal, and the movement of the mode of these distribution in phenotype space tracks the trajectory of the corresponding adaptive models. This is illustrated in the right panels of Figures 1 and 2.

\begin{figure*}
\centering
\includegraphics[width=0.3\textwidth]{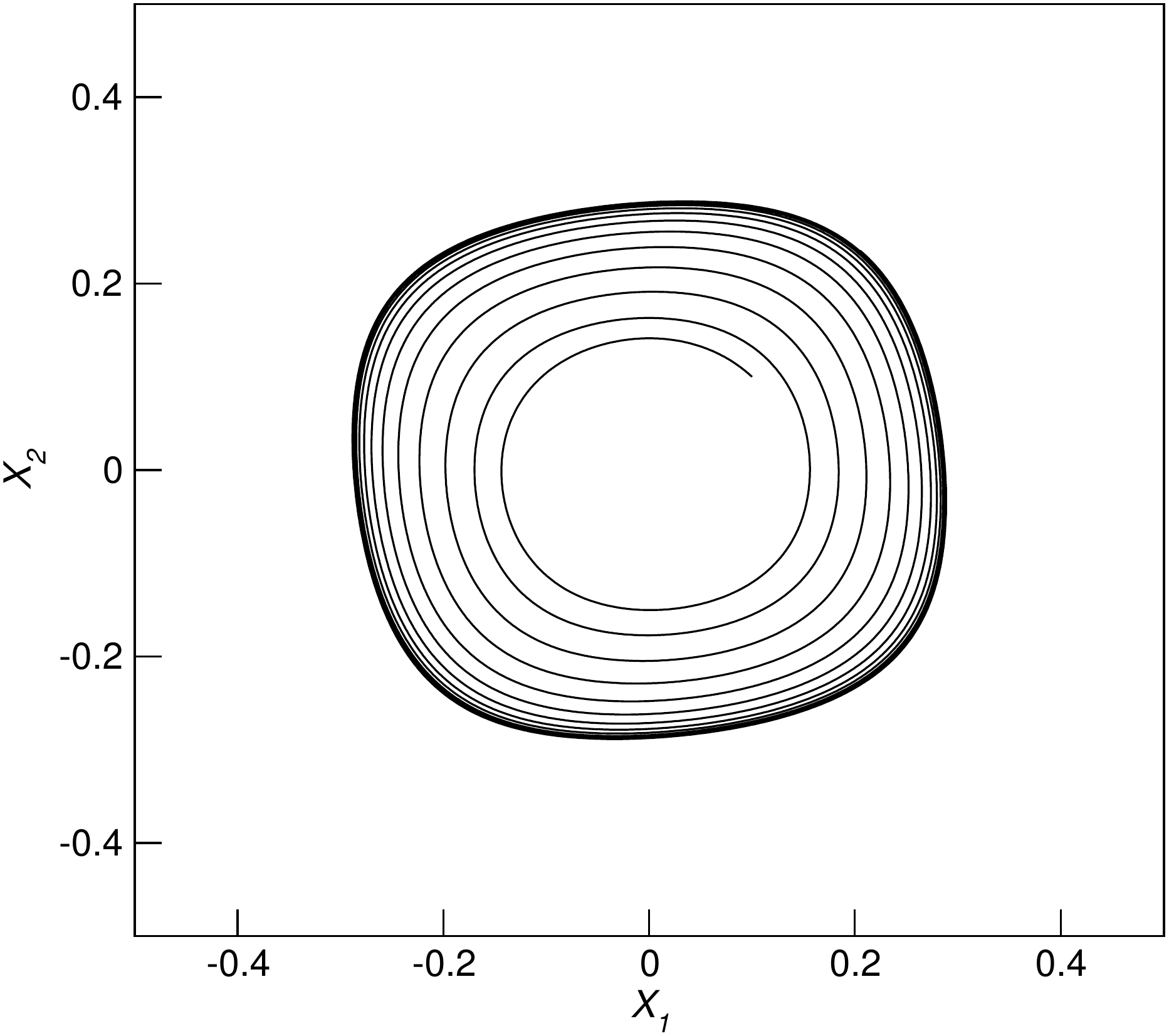}
\includegraphics[width=0.3\textwidth]{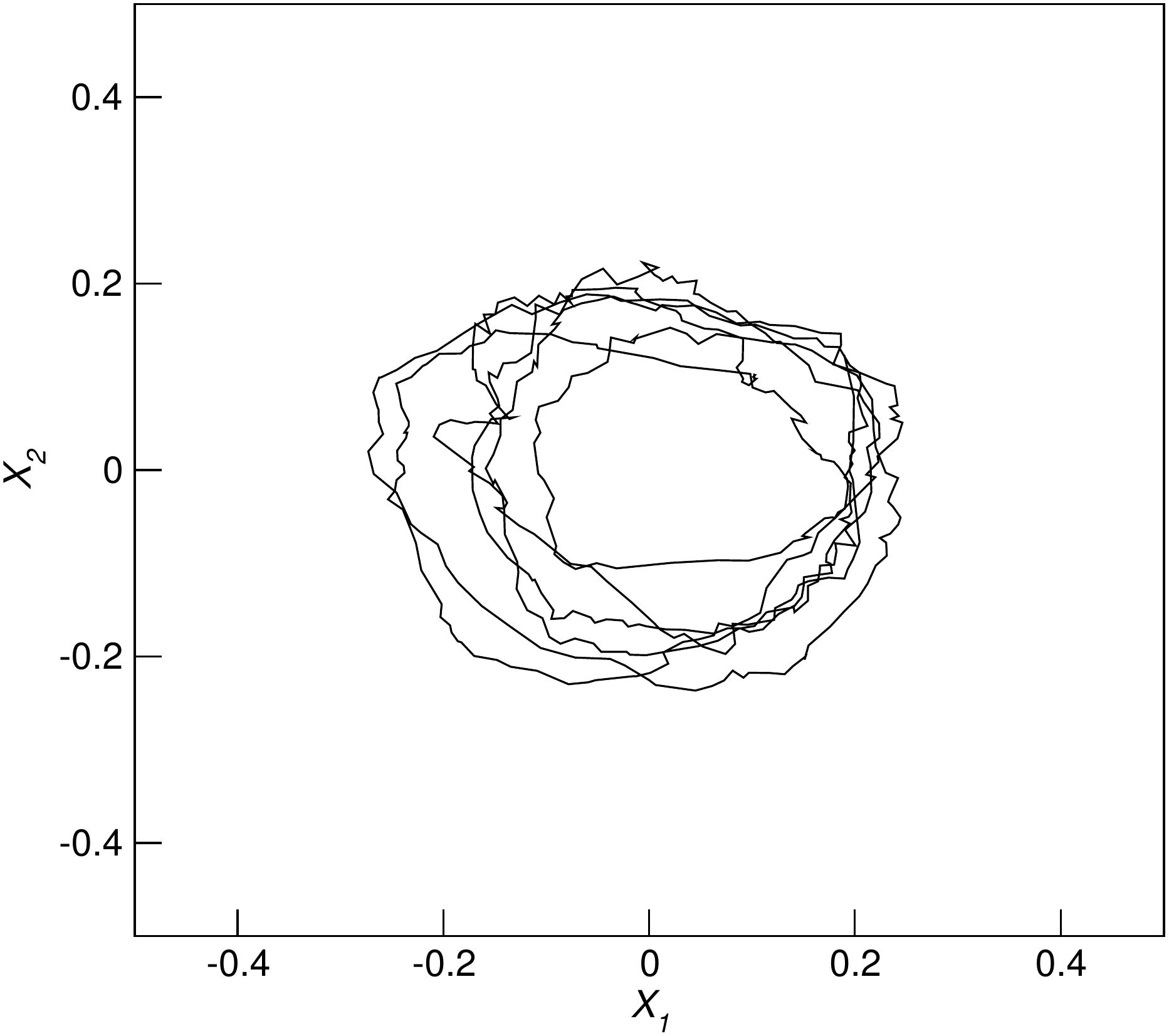}
\includegraphics[width=0.3\textwidth]{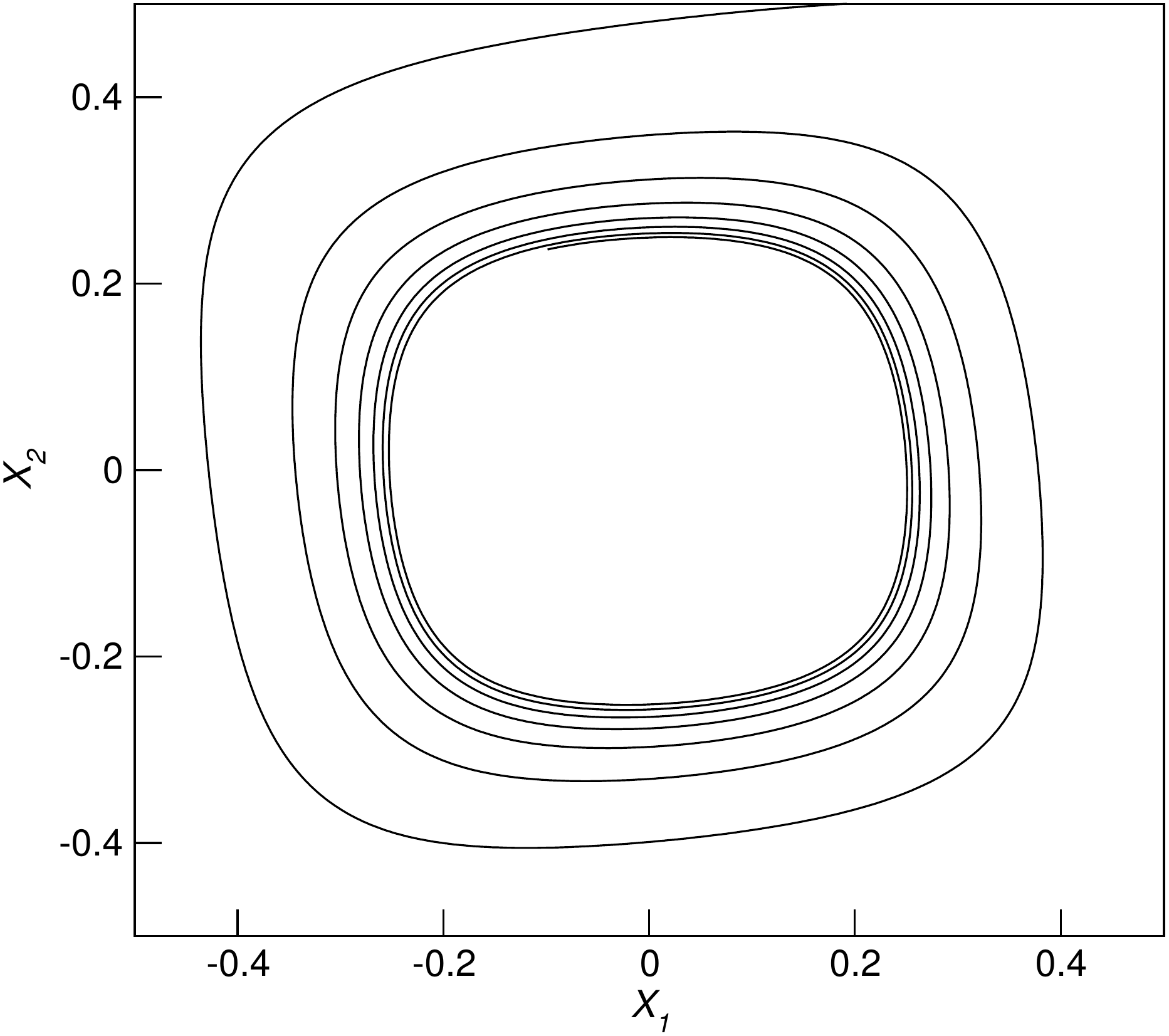}
\includemovie[poster]{5cm}{5cm}{video1.mov}
\includemovie[poster]{5cm}{5cm}{video4.mov}
\caption{An example of periodic adaptive dynamics  in a 2-dimensional phenotype space (left panel), the trajectory of the center of mass of the corresponding minimal individual-based model (center panel 
and the first video (online only) in the second row 
), and the trajectory of the center of mass of the corresponding minimal PDE model (right panel 
and the second video (online only) in the second row 
) for a 2-dimensional system. Projections  of the trajectories onto the first two phenotypic components are shown. The coefficients in the competition kernel (\ref{CCC}) are given in the file ``coeff2.dat'' in SOM.}
\label{f1}
\end{figure*}

\begin{figure*}
\centering
\includegraphics[width=0.3\textwidth]{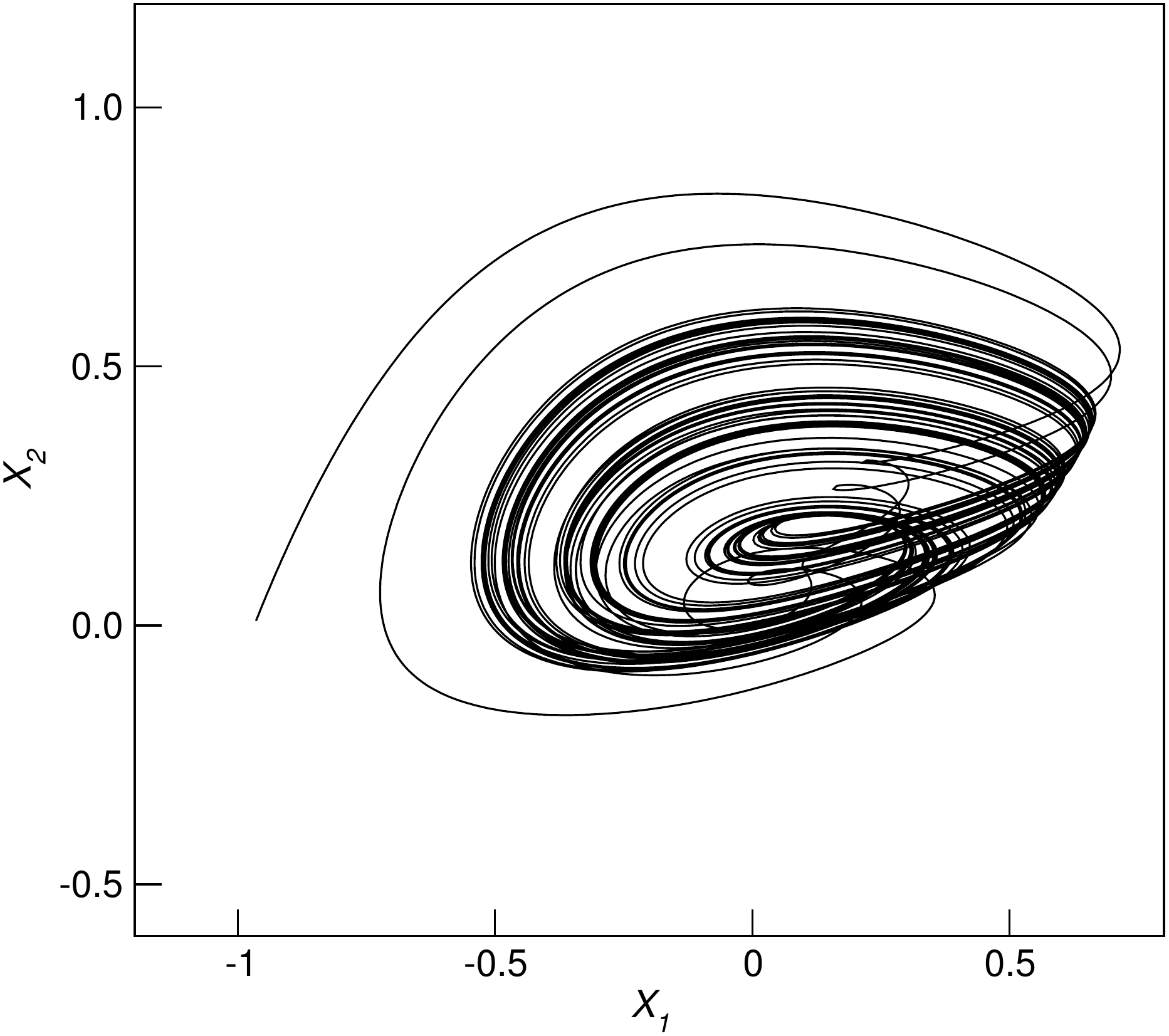}
\includegraphics[width=0.3\textwidth]{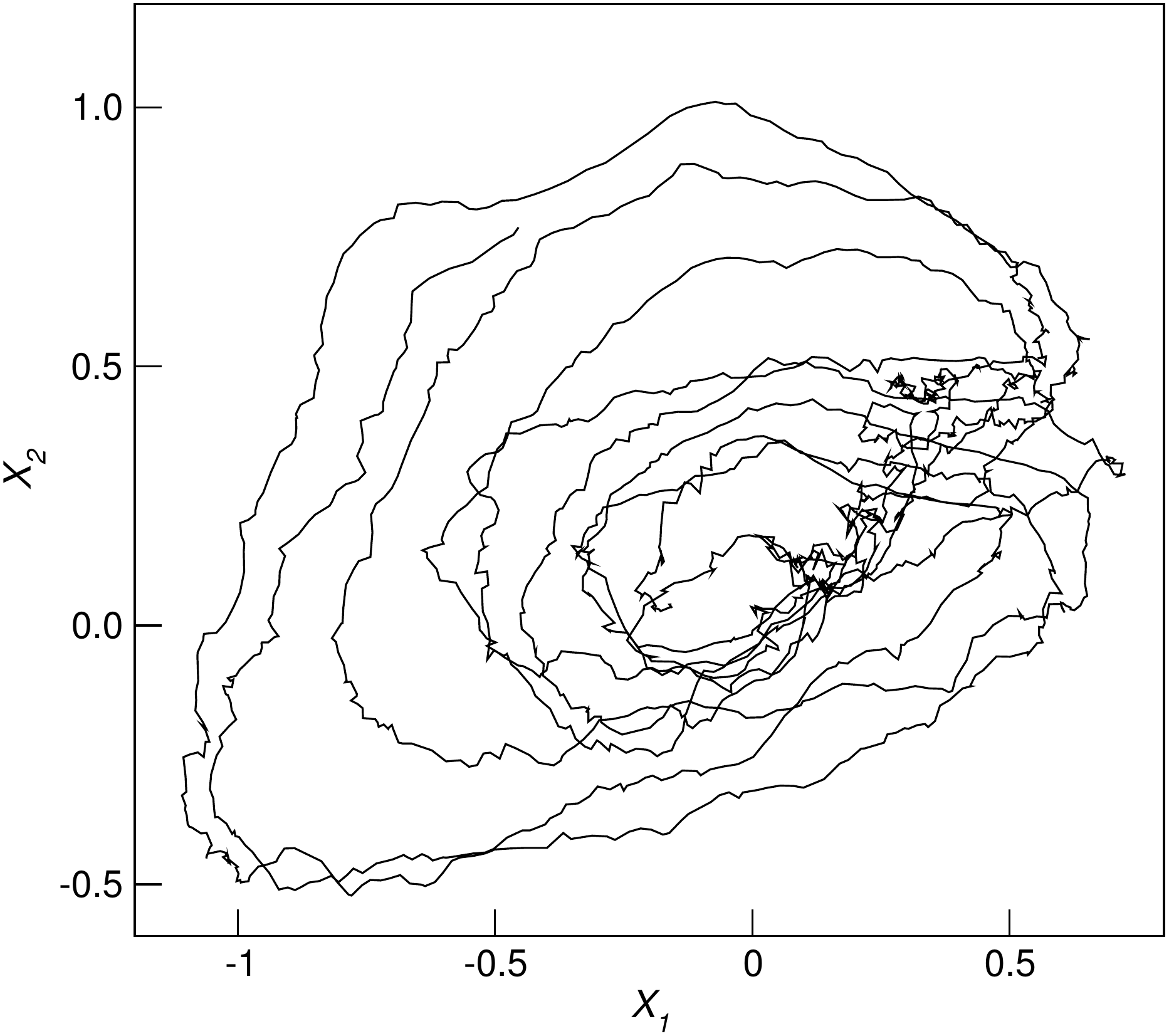}
\includegraphics[width=0.3\textwidth]{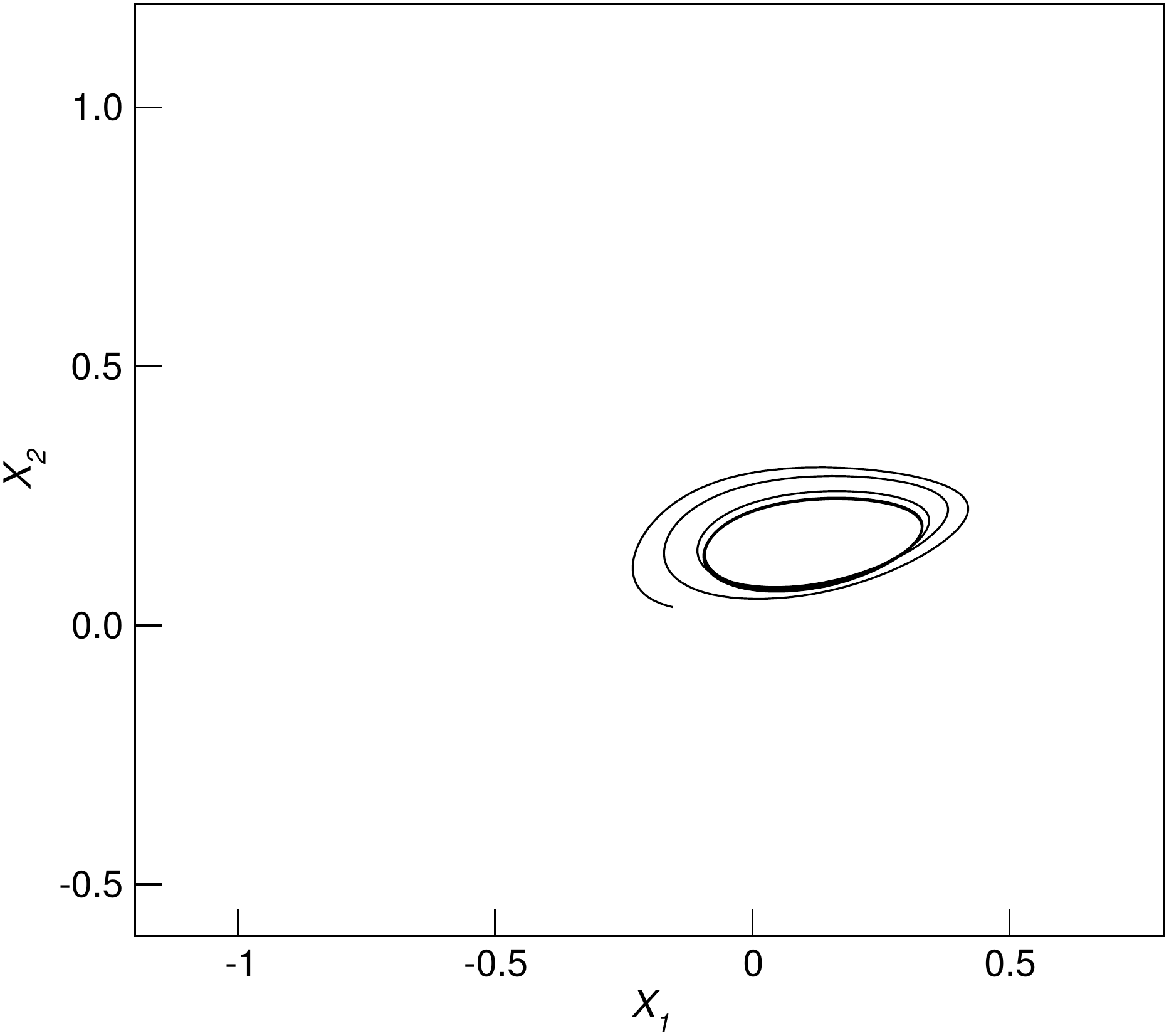}
\includemovie[poster]{5cm}{5cm}{video2.mov}
\includemovie[poster]{5cm}{5cm}{video5.mov}
\caption{An example of chaotic adaptive dynamics  in a 3-dimensional phenotype space (left panel), the trajectory of the center of mass of the corresponding minimal individual-based model (center panel
and the first video (online only) in the second row 
), and the trajectory of the center of mass of the corresponding minimal PDE model (right panel, 
and the second video (online only)  in the second row, 
only a small part of the attractor is reproduced since the PDE integration is very computationally extensive).  Projections of the trajectories onto the first two phenotypic components are shown. The coefficients in the competition kernel (\ref{CCC}) are given in the file ``coeff3.dat'' in SOM.}
\label{f2}
\end{figure*}

\begin{figure*}
\centering
\includegraphics[width=0.4\textwidth]{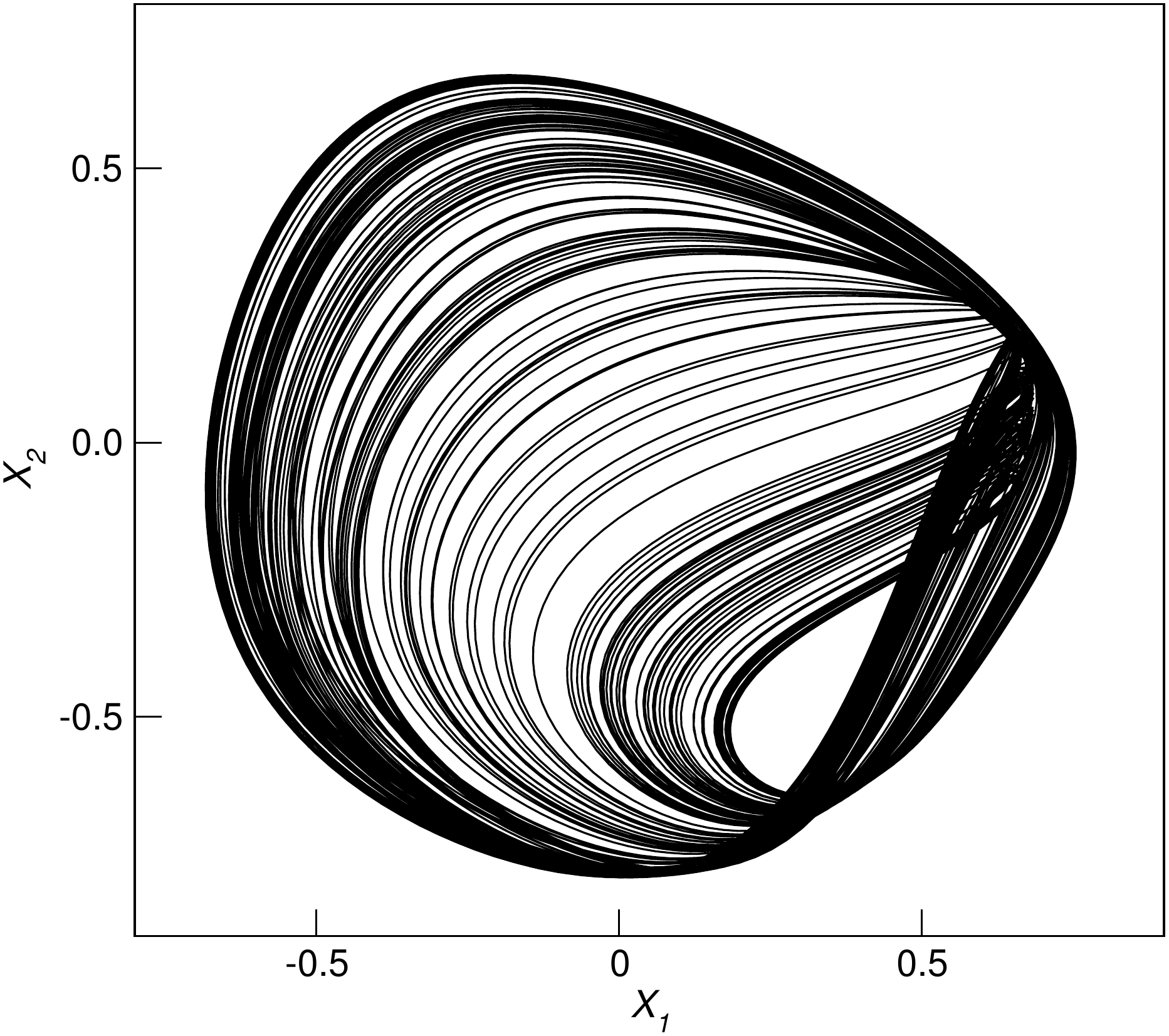}
\includegraphics[width=0.4\textwidth]{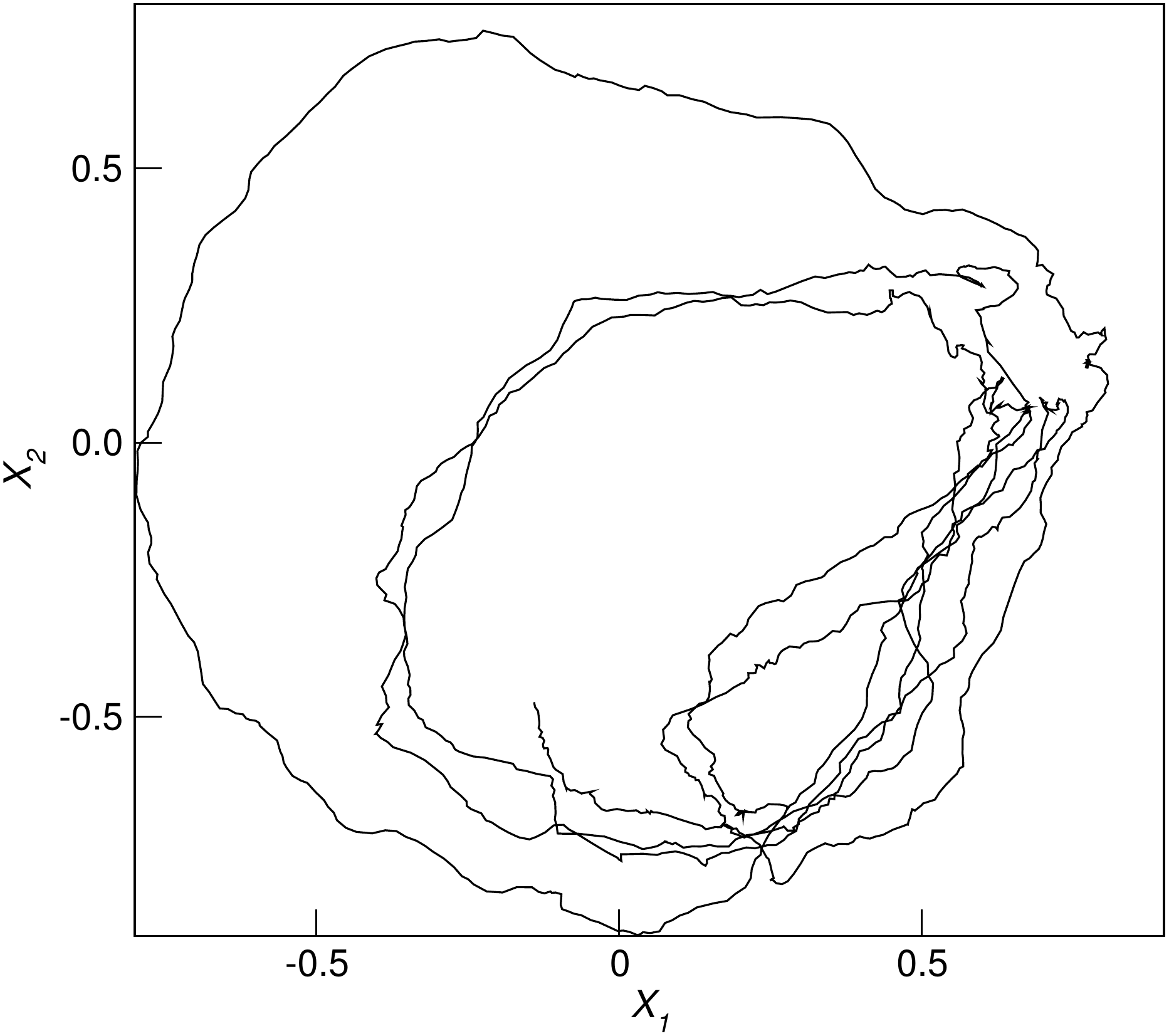}
\includemovie[poster]{5cm}{5cm}{video3.mov}
\caption{An example of chaotic adaptive dynamics in a 4-dimensional phenotype space (left panel) , the trajectory of the center of mass of the corresponding minimal individual-based model (right panel
and the video (online only) in the second row 
). Projections of the trajectories onto the first two phenotypic components are shown. The coefficients in the competition kernel (\ref{CCC}) are given in the file ``coeff4.dat'' in SOM.}
\label{f3}
\end{figure*}

\begin{figure}[ht]
\end{figure}

\subsection{Diversification with Gaussian competition kernels}
As mentioned before, the choice of  competition kernel (\ref{ACCC}) for the given adaptive dynamics (\ref{Gen_AD}) is by far not unique:
Specifically, consider the set of competition kernels of the form
\begin{align}
 \label{TCCC}
\alpha({\mathbf{x}, \mathbf{y}}) = &\exp\left[\sum_{i,j,k}^{d} (A_{ijk} x_ix_jx_k + B_{ijk}y_{i} x_jx_k + C_{ijk}    y_iy_jx_k\right. + D_{ijk}  y_iy_jy_k +E_{ij}x_ix_j + F_{ij} y_ix_j + G_{ij}y_i y_j)\Bigg],
\end{align}
where the sets of coefficients $\{A_{ink}\}$, $\{B_{ink}\}$, $\{C_{ink}\}$, $\{D_{ink}\}$, $\{E_{ij}\}$, $\{F_{ij}\}$, and $\{G_{ij}\}$ are subject to the following constraints:
\begin{align}
A_{ijk} + B_{ijk} + C_{ijk} + D_{ijk} &=0 \quad\text{for all}\quad i,j,k\nonumber\\
E_{ij} + F_{ij} + G_{ij}&=0 \quad\text{for all}\quad i,j\nonumber
\end{align}
and 
\begin{align}
a_{ijk} &= -(B_{ijk} + (C_{ijk} + C_{jik}) + (D_{ijk}+D_{jik}+D_{jki}))\nonumber\\
b_{ij} &= -(F_{ij} + (G_{ij}+G_{ji})),\nonumber
\end{align}
where $a_{ijk}$ and $b_{ij}$ are the coefficients determining the adaptive dynamics (\ref{main1}).
Then the first two of the above constraints ensure that $\alpha(\mathbf x, \mathbf x)=1$, and the third set of constrains ensures that the competition kernel (\ref{TCCC}), together with the carrying capacity (\ref{K}), generates the adaptive dynamics (\ref{main1}).

It is clear from these considerations that there are very many different competition kernels that give rise to the same adaptive dynamics. (In fact, the dimension of the space of order-3 competition kernels giving rise to one and the same adaptive dynamics model is $2d^3+d^2$.)

Here we are interested in those competition kernels corresponding to a given adaptive dynamics that would give rise to adaptive diversification. It has long been suggested that a general mechanism  to generate and maintain diversity is for competition to be strongest between similar phenotypes \cite{dieckmann_doebeli1999,doebeli2011, leimar_etal2013}. This is typically described by Gaussian competition kernels, which reflect the biologically plausible assumption that the strength of competition between individuals decreases with with phenotypic distance. 

In the present context, and starting with a given adaptive dynamics (\ref{main1}), we can incorporate Gaussian competition by multiplying the minimal competition kernel (\ref{ACCC}) with a Gaussian term, resulting in a competition kernel of the form 
\begin{align}
 \label{AG}
\alpha({\mathbf{x}, \mathbf{y}}) = \exp\left[\sum_{i=1}^{d} w_i(\mathbf{x})(x_i - y_i) - \frac{(x_i - y_i)^2}{2 \sigma_i^2}\right].
\end{align}

Note that this competition kernel is a particular case of the general form (\ref{TCCC}) (with $\{D\}=0$ and a suitable choice of coefficients $\{A\},\{B\},\{C\},\{E\},\{F\}$, and $\{G\}$). Again, together with the carrying capacity (\ref{K}), this competition kernel results in the general adaptive dynamics (\ref{main1}). But now the Hessian of the invasion fitness function has
an additional positive diagonal component given by the $\s_{i}^{-2}$,
\begin{align}
 \label{HG}
\mathbf{H}(\mathbf{x}) = &- \begin{pmatrix}  w_1 - u_1 \\...  \\w_d - u_d\end{pmatrix} \begin{pmatrix}  w_1 - u_1, & ...,& w_d - u_d\end{pmatrix} +\begin{pmatrix} \sigma_1^{-2}&...&0\\...&...&...\\0&...&\sigma_d^{-2}\end{pmatrix}+ \frac{\mathbf{H_K}(\mathbf{x})}{K(\mathbf{x})}.
\end{align}

This positive definite diagonal component is independent of the current resident phenotype, and hence in any phenotype space with dimension $d\geq2$, there are directions in phenotype space orthogonal to the current selection gradient along which there is a disruptive component of selection. For sufficiently small variances  $\sigma_k$  of the Gaussian components, there will therefore be a tendency towards diversification. For any given point on an evolutionary trajectory, i.e., for any given resident phenotype, the conditions for diversification are of course not only determined by the positive definite component of the Hessian, but also by the remaining terms in (\ref{HG}) and the selection gradient (\ref{Gen_AD}). However, if the  $\sigma_k$  of the Gaussian components are sufficiently small, the diversification conditions are likely to persist for a sufficient amount of time to make diversification possible along non-equilibrium evolutionary trajectories. 

Whether evolutionary branching can occur under these conditions can be checked by using the adaptive dynamics (\ref{main1}) from \cite{doebeli_ispolatov2014} as a starting point and constructing the corresponding individual-based model using the the competition kernel (\ref{AG}) and the carrying capacity (\ref{K}). 
For simplicity, we assumed that the variance of the Gaussian component $\sigma_k=\sigma$ was the same in all phenotypic directions, and our extensive simulations indicate that adaptive diversification indeed occurs whenever $\sigma$ is small enough. Figure 4 illustrates how  the typical interparticle separation, quantified as the ``total standard deviation'' $\s_{total}\equiv \sqrt{\sum_{i=1}^d \langle (x_i-\langle x_i\rangle )^2\rangle }$, becomes noticeably larger for  the individual-based model with the Gaussian competition kernel (\ref{AG}) compared to the same model without the Gaussian term.  The diversification in the models defined in Figs.~1-3 but with the Gaussian competition kernel  is further illustrated by videos in Figs.~5-7. The corresponding PDE solutions also exhibit diversification, which is shown in videos in the right panels in Figs.~5,6.

\begin{figure*}
\centering
\includegraphics[width=1.\textwidth]{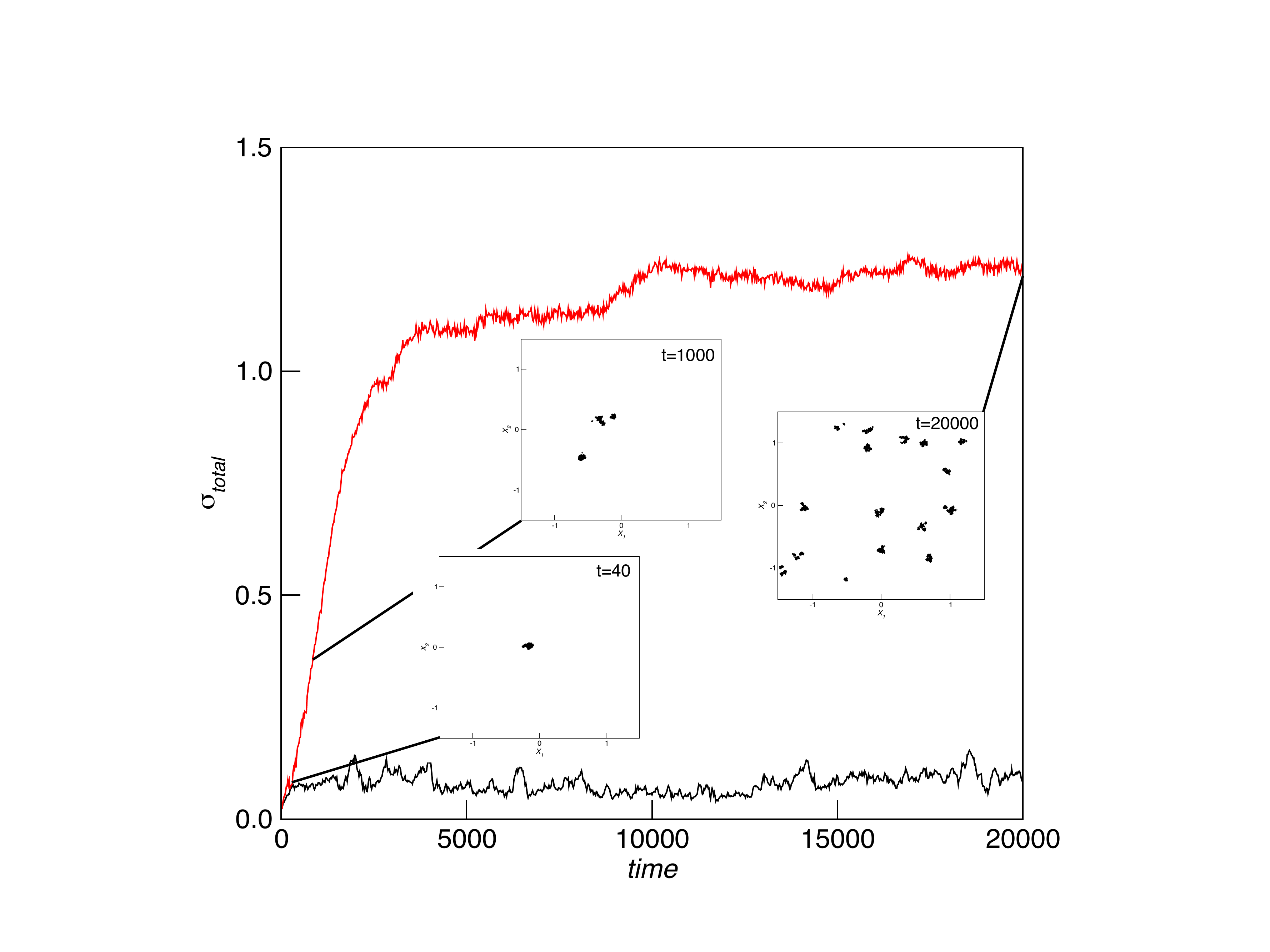}
\caption{Time dependence of the total standard deviation of the particles' coordinates defined as
$\s_{total}\equiv \sqrt{\sum_{i=1}^d \langle (x_i-\langle x_i\rangle )^2\rangle }$ for the 3-dimensional chaotic system as in Fig.~2 (bottom line), and with a Gaussian competition term (\ref{AG}) with the width $\s=0.65$ (top line, red online). Here $\langle\ldots\rangle$ stands for an average over all particles present in the system.  Three snapshots for the Gaussian competition kerne illustrate the initial, single-cluster  stage at $t=40$, intermediate diversification at $t=1000$, and the well-developed steady-state diversification at $t=20 000$ are shown in inserts. The time evolution that has led to these snapshots can be seen in  video in Fig.~6.}
\label{f4}
\end{figure*}  

\begin{figure*}
\centering
\includemovie[poster]{5cm}{5cm}{video6.mov}
\includemovie[poster]{5cm}{5cm}{video9.mov}
\caption{(Online video) Diversification of the individual-based model (left panel) and the PDE solution (right panel) of the 2-dimensional periodic system presented in Fig.~1 but with a Gaussian competition term (\ref{AG}) with the width $\s=0.65$.}
\label{f5}
\end{figure*}
\begin{figure*}
\centering
\includemovie[poster]{5cm}{5cm}{video7.mov}
\includemovie[poster]{5cm}{5cm}{video10.mov}
\caption{(Online video) Diversification of the individual-based model (left panel) and the PDE solution (right panel) of the 3-dimensional chaotic system presented in Fig.~2 but with a Gaussian competition term (\ref{AG})  with the width $\s=0.65$.}
\label{f6}
\end{figure*}
\begin{figure*}
\centering
\includemovie[poster]{5cm}{5cm}{video8.mov}
\caption{(Online video) Diversification of the individual-based model  of the 4-dimensional chaotic system presented in Fig.~3 but with a Gaussian competition term (\ref{AG})  with the width $\s=0.5$.}
\label{f7}
\end{figure*}

As mentioned, the Gaussian competition kernel (\ref{AG}) is a special case of the general competition kernel (\ref{TCCC}), and there are potentially many more sets of coefficients in (\ref{TCCC}) that generate the same adaptive dynamics (\ref{main1}) but lead to diversification in the corresponding individual-based models. Our goal here was not to enumerate all forms of $\alpha$ that can produce evolutionary branching, but rather to show that a particular form of competition kernel generates models that do not diversify and follow the adaptive dynamics trajectories, while competition kernels incorporating sufficiently strong frequency-dependence leads to robust and persistent diversification in systems with non-stationary adaptive dynamics. 





\section{Conclusions} 
 
Recent advances in understanding the evolution of complex multidimensional phenotypes were mainly limited to the adaptive dynamics framework \cite{ito_dieckmann2007,doebeli_ispolatov2010,doebeli_ispolatov2014,svardal_etal2014} or to the the quantitative genetics framework \cite{gilman_etal2012, debarre_etal2014}. Neither of these frameworks lends itself directly to the study of the evolutionary dynamics of diversification and subsequent co-evolution of coexisting phenotypic clusters, as adaptive dynamics assumes monomorphic populations, while quantitative genetics models typically assume Gaussian phenotypic distributions.  Here we have made an attempt to go beyond these approximations by studying stochastic, individual-based model as well as deterministic PDE models for adaptive diversification under non-equilibrium evolutionary dynamics. We considered a general class of models for evolution due to frequency-dependent competition. Starting with a given adaptive dynamics model in $d$-dimensional phenotype space ($d\geq1$) with potentially complicated dynamics, we constructed individual-based models that give rise to this adaptive dynamics model in the limit of large populations, and rare and small mutations. Without the assumption of small and rare mutations, the individual-based models give rise to deterministic PDE models in the large population limit \cite{champagnat_etal2006, champagnat_etal2008}, which in turn give rise to the given adaptive dynamics model in the limit of small variance of the evolving phenotypic distributions. 

Our analysis generated two basic conclusions:
\begin{itemize}
\item For a given adaptive dynamics in multidimensional phenotype space there is a corresponding  ``minimal'' individual-based model whose fitness Hessian is always negative definite; as a consequence, this minimal individual-based model does not diversify, and instead the population remains confined to a single cluster whose centre of mass follows the trajectory of the corresponding  adaptive dynamics model, regardless of the nature of the attractor of the adaptive dynamics.
\item Multiplying the minimal competition kernel by a Gaussian term with a sufficiently  small width yields the same adaptive dynamics, but causes adaptive diversification in the individual-based model, in which multiple coexisting clusters emerge. Again this holds regardless of the nature of the attractor of the given adaptive dynamics. In particular, adaptive diversification is possible from a complicated evolutionary trajectory.
\end{itemize}

Similar statements hold for the corresponding PDE models. In particular, diversification is possible in the absence of evolutionary equilibrium attractors in phenotype space, and in high-dimensional phenotype spaces diversification can in principle occur in any direction that is orthogonal to the selection gradient. These findings considerably widen the scope of adaptive diversification as a general evolutionary principle. We have also shown that the correspondence between individual;-based and PDE models on the one hand, and adaptive dynamics models on the other hand, is not unique: there are very many different individual-based and PDE models that give rise to the same  canonical equation \cite{dieckmann_law1996} for the monomorphic adaptive dynamics. According to the analysis in \cite{ito_dieckmann2014}, diversification along complicated trajectories in these models should depend on the components of the Hessian matrix of second derivatives of the invasion fitness function that are orthogonal to the selection gradients, and our results corroborate this. In particular, different ``full'' models reconstructed from a given adaptive dynamics  have different orthogonal Hessians and different diversification properties. At one end of this set of full models are the minimal models, which generate a negative definite orthogonal Hessian, and therefore do not show diversification. The other extreme is the Gaussian model, which can be defined to have Hessians with positive eigenvalues in all orthogonal directions, regardless of the current resident phenotype. Consequently, these reconstructed ``Gaussian'' models have a high propensity to diversify. 

It is interesting to note that the minimal models  reproduce the adaptive dynamics attractor even if this attractor is chaotic. Due to the intrinsic sensitivity and complexity of chaotic systems, it was not clear a priori that this is possible. For the Gaussian models presented here we made the simplifying assumption that competition between similar phenotypes is equally strong in all phenotypic directions. In reality, Gaussian competition could act only in a subset of phenotypic directions, and then one would expect diversification to primarily occur in this subset. Also, we assumed all off-diagonal elements in the Hessian portion of the competition kernel to be 0, which corresponds to assuming that there are no interactions between the phenotypic components with regard to the Gaussian component of competition. It is known that such interactions promote diversification in equilibrium models \cite{doebeli_ispolatov2010, svardal_etal2014, debarre_etal2014}, and it is an interesting direction for future work to analyze whether similar results hold for diversification in non-equilibrium models. 

Our simulation results indicate that diversification often leads to coexistence of multiple phenotypic  clusters, and a very interesting question for future work concerns the effect of diversification on the complexity of the co-evolutionary dynamics of these coexisting clusters. So far, we have seen examples of both destabilization and stabilization due to adaptive diversification: populations converging to an equilibrium when monomorphic can embark on non-equilibrium dynamics after diversification; and  the opposite can happen as well, so that populations moving on a complicated attractor when monomorphic converge to a multi-cluster evolutionary equilibrium after diversification.

It seems clear that in general, many different phenotypic properties can contribute to ecological interactions, resulting in potentially complicated selection pressures in high-dimensional phenotype spaces. Our previous work  has highlighted both the increased propensity for evolutionary branching in such spaces \cite{doebeli_ispolatov2010}, and the increased propensity towards complicated evolutionary trajectories \cite{doebeli_ispolatov2014}. Here we have extended this perspective to studying the full dynamics of diversification in high-dimensional phenotype spaces  in stochastic, individual-based models as well as PDE models. When these evolving systems diversify, the resulting coevolutionary dynamics of coexisting phenotypic clusters may often be even more complicated and unpredictable, thus further challenging the prevailing view of evolution as an optimization and equilibrium process. We hope that our work will contribute to the discussion about the importance of adaptive diversification as a mechanism for generating biological diversity. Finally, we think that studying individual-based and PDE models of diversification in high dimensional spaces can also be relevant for tackling important  questions in cultural evolution, such as the origin and evolution of different religions, languages, and other cultural traditions \cite{doebeli2011}.

\begin{acknowledgments}
I. I. was supported by FONDECYT (Chile) grants 1110288 and 1151524. M. D. was supported by NSERC (Canada). All authors contributed equally to this work. 
\end{acknowledgments}

\appendix
\section{Supplementary online materials}

Here we describe how we simulate the individual-based and PDE models introduced in the main text. Both models are defined by the competition
kernel 
\begin{align}
 \label{CK}
\alpha({\bx, \by}) = \exp\left[\sum_{j,j=1}^{d}
  b_{ij}x_{j}(x_i - y_i) + \sum_{i,j,k=1}^{d} a_{ijk} x_jx_k(x_i -
  y_i)+\right. 
\left.  \sum_{i=1}^{d} \frac{(x_i - y_i)^2}{2 \sigma_i^2} \right],
\end{align}
and the carrying capacity
\begin{align}
 \label{CC}
 K(\bx) = K_0\exp \left( -\sum_i^d x_i^4/4\right).
\end{align} 
Since we are mainly interested in non-equilibrium dynamics, the sets of
coefficients $\{a\}$ and $\{b\}$ and the initial conditions $\bx_0$
were selected as described in \cite{doebeli_ispolatov2014}, 
so that the corresponding adaptive dynamics, given by
Eq. 13 in the Main Text, was either cyclic, quasiperiodic or chaotic.

\subsection {Individual-based simulation}
Individual-based realizations of the model
were based on the Gillespie algorithm \cite{gillespie1976}
and consisted of the following steps:
\begin{enumerate}
\item  The system is initialized by creating a set of $K_0 \sim 10^3 - 10^4$ individuals with
  phenotypes  $\bx_{\a}\in\mathbf{R}^d$ localized around the initial position $\bx_0$
  with a small random spread $|\bx_{\a} - \bx_0|\sim10^{-3}$.
\item  Each individual $\a$ has a 
constant reproduction rate $\r_{\a}=1$ and a death rate $\d_{\a}=\sum_{\b\neq \a}
A(\bx_{\a},\bx{_\b})/K(\bx_{\a})$, as defined by logistic ecological dynamics.
\item  The total update rate is
given by the sum of all individual rates $U=\sum_{\a}
(\r_{\a}+\d_{\a})$. 
\item The running time $t$ is incremented by a random number
$\D t$ drawn from the exponential distribution $P(\D t)= U \exp (-\D t  U)$.
\item A particular birth or death event is randomly chosen  with
probability equal to the rate of this event divided by the total update
rate $U$. If  a reproduction event is chosen, the phenotype of an
offspring is offset from the ancestral one by a 
small mutation randomly drawn from a uniform distribution with
amplitude $\m = 10^{-3} - 10^{-2}$.
\item The individual death rates $\d_{\a}$ and the total update rate
$U$ are updated  to take into account the addition or removal of an
individual.
\item  Steps 4-6 are repeated until $t$ reaches a specified end time. 
\end {enumerate}

\subsection{Partial differential equation models}
A deterministic large-population limit of the individual-based model is obtained as the partial differential equation (PDE)
\begin{align}
\label{logistic}
 \frac{\partial N(\bx, t)}{\partial t} = N(\bx, t)\left(\frac{ 1 - \int \alpha(\by, \bx) N (\by, t) dy}{K(\bx)}\right)+D\sum_{i=1}^d \frac{\partial^2 N(\bx, t)}{\partial x_i^2},
\end{align}
where  $N(\bx, t)$ is the population distribution at time $t$ \cite{champagnat_etal2006}. The second term of the right hand side is a diffusion term that describes mutations, 
with the diffusion coefficient typically set to $D\sim 10^{-4} - 10^{-3}$.
The form and size of the single-cluster trajectory was usually known from
the adaptive dynamics solution. Hence, to numerically solve the PDE model (\ref{logistic})
 we chose a finite lattice resolution of phenotype space nodes at least twice
larger than the adaptive dynamics attractor. The number of bins $L$
in each dimension of this lattice is strongly limited by memory limitations: An
efficient implementation requires computing and storing an array of
$L^{d} \times L^d$ values of the competition kernel $\alpha(\by_i, \bx_j)$ for the pairwise interactions between
all sites $i$ and $j$.  With $L=25 -30$ to achieve a reasonable
spatial resolution, the memory constraint makes the
PDE implementation feasible only for $d=2,3$. 

\bibliography{EvolutionofDiversity}
\bibliographystyle{prslb}

\end{document}